\documentclass[aps,prd]{revtex4}
\usepackage{graphics}
\usepackage{psfrag}
\begin{document} 

\title{
Scalar perturbation spectra from warm inflation} 
\author{Lisa M. H. Hall}
\email{lisa.hall@ncl.ac.uk}
\author{Ian G. Moss}
\email{ian.moss@ncl.ac.uk}
\affiliation{School of Mathematics and Statistics, University of  
Newcastle Upon Tyne, NE1 7RU, UK}
\author{Arjun Berera} \email{ab@ph.ed.ac.uk}
\affiliation{School of Physics, University of Edinburgh,
Edinburgh, EH9 3JZ, U.K.}

\date{\today}


\begin{abstract}
We present a numerical integration of the cosmological scalar perturbation
equations in warm inflation. The initial conditions are provided by a
discussion of the
thermal fluctuations of an inflaton field and thermal radiation using a
combination of thermal field theory and thermodynamics. The perturbation
equations include the effects of a damping coefficient $\Gamma$ and a
thermodynamic potential $V$. We  give an analytic expression for the spectral
index of scalar fluctuations in terms of a new slow-roll parameter
constructed from $\Gamma$. A series of toy models, inspired by spontaneous
symmetry breaking and a known form of the damping coefficient, lead to a
spectrum with $n_s>1$ on large scales and $n_s<1$ on small scales.
\end{abstract}

\maketitle
\section{Introduction}

Inflation is the most successful idea which we have available
for explaining the large scale structure of the universe.
Inflationary dynamics can be realized in two distinct
ways. In the original picture, termed supercooled inflation, the universe
rapidly supercools during an inflation phase and subsequently a reheating
period is invoked to end inflation and fill the universe with radiation
\cite{guth81,linde82,albrecht82,hawking82}. In
the other picture, termed warm inflation, dissipative effects are important
during the inflation period, so that radiation production occurs concurrently
with inflationary expansion.

The idea of warm inflation was influenced by a calculation of the friction
term in the inflaton equation of motion by Hosoya and Sakagami \cite{hosoya84}.
The magnitude of the damping term suggested the possibility that it could be
the dominant effect prolonging inflation
\cite{moss85,lonsdale87,yokoyama88,liddle89}. The significance of
dissipation to inflation was independently realized in \cite{berera95},
but this time with a clearer picture of the associated inflation dynamics, and
named warm inflation.  This and associated works \cite{berera96,berera97} also
outlined the nonequilibrium thermodynamical problem underlying this picture,
which has subsquently has been studied in greater detail
\cite{berera98,berera99,yokoyama99,moss02,berera01,berera03,lawrie01}.
These works confirmed that at high temperature the damping
was proportional to the relaxation time of the radiation.
In addition to these studies of warm inflation dynamics,
several phenomenological warm inflation models have been discussed
in the literature 
\cite{oliveira98,bellini98,maia99,chimento02,oliveira02-2}.

The density fluctuations in warm inflation arise from thermal, rather than
vacuum, fluctuations \cite{moss85,bererafang95,berera00}. These have their
origin in the hot radiation and influence the inflaton through a noise term in
the equations of motion \cite{berera96,berera00}.

The development of cosmological perturbations in warm inflation has been
investigated by several authors. Taylor and Berera \cite{taylor00} have
analysed the perturbation spectra by matching the thermally produced
fluctuations to gauge invariant parameters when the fluctuations cross the
horizon. This technique should be good for order of magnitude estimates. 
For more accurate treatment of the density perturbations it is nessesary to use
the cosmological perturbation equations, which can be found in the literature
\cite{lee00,oliveira01,hwang02}.

The main focus of the present work is to solve the perturbation equations
numerically. We will also set up the equations to take into account some
effects which have previously been ignored. We introduce a new slow-roll
parameter and include modifications to the equations when the damping term and
the potential depend on temperature.

Other improvements have been made in the treatment of thermal fluctuations. The
influence of the cosmological expansion on the inflaton fluctuations is
considered in detail. This gives a clearer and more accurate picture of the
development of the thermal fluctuations than previously. Thermal fluctuations
in the radiation, which lead to entropy perturbations, are also analysed.
Entropy perturbations are always present during warm inflation and react back
on the curvature fluctuations. In the basic model, the entropy fluctuations
disappear before inflation ends.

Section 2 contains an updated outline of the theory of warm inflation. In
section 3 we obtain expressions for the thermal fluctuations of the inflaton
and the radiation. Cosmological perturbations are discussed in section 4. The
results of numerical solutions for the cosmological perturbations using the
thermal fluctuations as initial conditions are compared with observations in
section 5. A summary of the main results and some general comments appear in
section 6.    
\section{Warm Inflation}

In this section we shall consider the case of a homogeneous inflaton field
interacting with thermalised radiation. If the inflaton is evolving very
slowly, then the radiation creates a thermal correction to the inflaton
potential \cite{dolan74} and a damping force \cite{hosoya84}. The system of
equations describing the system include an equation for the inflaton and an
equation describing how the energy lost by the inflaton through the damping
force is transfered to the radiation.

The equation describing the effects of radiation damping on the evolution of
the inflaton field was first obtained by Hosoya \cite{hosoya84} using a form
of thermodynamic transport theory. This is applicable as long as the inflaton
evolves slowly and the radiation remains close to thermal equilibrium. More
recent treatments are based on the closed time path formalism, and an account
of the effects of radiation damping on an evolving field in flat space can be
found, for example, in a paper by Gleiser and Ramos \cite{gleiser94}. An
interesting variation occurs when the inflaton interacts with the radiation
via an intermediate particle decay \cite{berera03}.  

When the early universe is in homogeneous expansion with expansion rate $H(t)$,
the evolution equation for the inflaton field is given by
\begin{equation}
\ddot\phi+(3H+\Gamma)\dot\phi+V_{,\phi}=0,\label{wip}
\end{equation}
where $\Gamma(\phi,T)$ is the damping term and $V(\phi,T)$ is the
thermodynamic potential. The damping term has a generic form given
approximately by $\Gamma\sim g^4\phi^2\tau$, where $g$ is a coupling constant.
For Hosoya's damping term, $\tau\equiv\tau(\phi,T)$ is related to the
relaxation time of the radiation
and for the models with an intermediate particle decay $\tau\equiv\tau(\phi)$
is related to the lifetime of the intermediate particle. The thermodynamic
potential can be expressed in a standardised way by
\cite{linde79}
\begin{equation}
V(\phi,T)=-{\pi^2\over 90}g_*T^4+{1\over 2}(\delta m_T)^2\phi^2
+V_0(\phi),\label{edens}
\end{equation}
where $g_*(T)$ is the effective particle number and $\delta m_T(\phi,T)$
represents thermal corrections.

The relative strength of the thermal damping
compared to the expansion damping can be described by a parameter $r$,
\begin{equation}
r={\Gamma\over 3H}
\end{equation}
Warm inflation is defined to be inflation with large values of $r$, as opposed
to supercooled inflation in which $r$ can be neglected.    

The dissipation of the inflaton's motion is associated with the production of
entropy. The entropy density of the radiation $s(\phi,T)$ is defined by
a thermodynamic relation in terms of the thermodynamic potential,
\begin{equation}
s=-V_{,T}.
\end{equation}
The rate of entropy production can be deduced from the conservation of
energy-momentum. The total density $\rho$ and pressure $p$ are given by,
\begin{eqnarray}
\rho&=&{\textstyle\frac12}\dot\phi^2+V+Ts\label{density}\\
p&=&{\textstyle\frac12}\dot\phi^2-V.
\end{eqnarray}
Energy-momentum conservation,
\begin{equation}
\dot\rho+3H(p+\rho)=0,
\end{equation}
now implies entropy production. Making use of Eq. (\ref{wip}) we get
\begin{equation}
T(\dot s+3Hs)=\Gamma\dot\phi^2.\label{wis}
\end{equation}
The zero curvature Friedman equation completes the set of differential
equations for $\phi$, $T$ and the scale factor $a$,
\begin{equation}
3H^2=8\pi G({\textstyle\frac12}\dot\phi^2+V+Ts).\label{wia}
\end{equation}

The entropy production has been described in a slightly different way in some
previous work on warm inflation \cite{berera95,berera00}. We can recover an
alternative equation in the case when the temperature corrections to the
potential are negligeable. If we set $\delta m_T=0$ in Eq. (\ref{edens}), then
the
radiation density $\rho_r=4sT/3$, and Eq. (\ref{wis}) becomes
\begin{equation}
\dot \rho_r+4H\rho_r=\Gamma\dot\phi^2.\label{wir}
\end{equation}
This equation is {\it only} valid when $\delta m_T=0$.
 
The slow-roll approximation consists of neglecting terms
in the preceeding equations with the highest order in time derivatives  
\begin{eqnarray}
\dot\phi&=&-{V_{,\phi}\over 3H(1+r )}\label{slowp}\\
Ts&=&r \dot\phi^2\label{slows}\\
3H^2&=&8\pi GV.\label{slowh}
\end{eqnarray}
Note that $Ts$ is the same order as two time derivatives. Slow-roll
automaticaly implies inflation, $\ddot a>0$.

The consistency of the slow-roll approximation is governed
by a set of slow-roll parameters. Warm inflation has extra slow-roll
parameters in addition to the usual set for supercooled inflation due to the
presence
of the damping term $\Gamma$. There are four `leading order' slow-roll
parameters,
\begin{equation}
\epsilon={1\over 16\pi G}\left({V_{,\phi}\over V}\right)^2,\quad
\eta={1\over 8\pi G}\left({V_{,\phi\phi}\over V}\right),\quad
\beta={1\over 8\pi G}\left({\Gamma_{,\phi}V_{,\phi}\over \Gamma V}\right),\quad
\delta={TV_{,\phi T}\over V_{,\phi}}.
\label{slowrp}
\end{equation}
The first two parameters are the standard ones introduced for supercooled
inflation \cite{liddle94,stewart02,habib02,martin03}. Two new parameters are
required for the $\phi$ dependence of the damping term and the temperature
dependence of the potential. The slow-roll approximation is valid when all of
the slow-roll parameters are smaller than $1+r$. In supercooled inflation, the
condition is tighter and the slow roll parameters have to be smaller than 1.

As an example, consider the case $\Gamma_{,T}=V_{,\phi T}=0$. In the warm
inflationary regime $r\gg 1$, the rate of change of various slowly varying
parameters are given by differentiating equations (\ref{slowp}-\ref{slowh}),
\begin{eqnarray}
{1\over H}{d\ln H\over dt}&=&-{1\over r}\epsilon,\label{hdot}\\
{1\over H}{d\ln \dot\phi\over dt}&=&-{1\over r}(\eta-\beta),\\
{1\over H}{d\ln Ts \over dt}&=&-{1\over r}(2\eta-\beta-\epsilon).\label{tdot}
\end{eqnarray}
In each case, the right hand size of these equations gives the relative sizes
of terms neglected in the slow-roll approximation. Slow-roll therefore requires
$\epsilon\ll r$, $\eta\ll r$ and $\beta\ll r$.

Inflation ends at the time when $\ddot a=\dot H+H^2$ falls to $0$. From
Eq. (\ref{hdot}), the end of inflation corresponds to the condition
$\epsilon=r$.
Furthermore, the slow-roll equations allow us to rewrite this condition as
$\rho_r=V$, in other words inflation ends when the radiation energy density
reaches the same value as the vacuum energy.

Similar arguments apply when $\Gamma_{,T}$ and $V_{,\phi T}$ are non-zero, for
example,
\begin{equation}
{1\over H}{d\ln \dot\phi\over dt}=-{1\over r}
\left({4-c\over 4+c}\eta-{4\over 4+c}\beta+{c\over 4+c}\epsilon-
{3c\over 4+c}\delta\right)
\end{equation}
Note that $c=T\Gamma_{,T}/\Gamma$ is not required to be small and therefore
does not enter the list of slow-roll parameters. The slow-roll approximation is
valid for $\epsilon\ll r$, $\eta\ll r$, $\beta\ll r$ and $\delta\ll r$.
  
\section{Thermal Fluctuations}

In this section we shall consider the behaviour of thermal fluctuations during
inflation and their influence on density perturbations on length scales smaller
than the size of the horizon. Thermal fluctuations, if they are present, form
the predominant source of density perturbations in warm inflation. We will
assume throughout that the radiation is close to thermal equilibrium.
Surprisingly, this is not a necessary condition for warm inflation but it is
an important regime. Conditions for thermalisation are model dependent, and
are discussed in refs \cite{berera98,berera99,berera01,berera03}.

The evolution of a comoving mode of the inflaton fluctuations during a warm
inflationary era can be divided into three regimes, depending on the relative
effect of different physical processes:
\begin{enumerate}
\item Thermal noise
\item Expansion
\item Curvature fluctuations,
\end{enumerate}
The transitions between these regimes are called freezeout and horizon
crossing. Where previous calculations have treated each of these regimes
seperately, we shall consider the combined effects of the thermal noise and
the expansion, and then the combined effects of expansion and curvature
fluctuations. The main results of this section are equations (\ref{dphi}),
(\ref{ddphi}) and (\ref{drho}), which provide the initial conditions for the
classical evolution equations described in the next section.

\subsection{Inflaton fluctuations}

The behaviour of a scalar field interacting with radiation can be analysed
using the Schwinger-Keldysh approach to non-equilibrium field
theory\cite{schwinger61,keldysh64}. According to this approach, a simple
picture emerges in which the field evolves by a Langevin equation
\cite{calzetta88,gleiser94},
\begin{equation}
-\nabla^2\phi(x,t)+\Gamma\dot\phi(x,t)+V_{,\phi}=\xi(x,t),
\end{equation}
where $\xi$ is a stochastic noise source. Correlation functions can be
evaluated by probability averages. 

The Schwinger-Keldeysh approach has been studied for curved, as well as flat,
spacetime backgrounds \cite{calzetta87}. Although the particular form of the
Langevin equation has not yet been fully investigated, we can appeal to the
equivalence principle to deduce the form of the Langevin equation on scales
smaller than the horizon \cite{berera00}.

We shall analyse the inflaton fluctuations on an expanding background over
timescales which are longer than the microphysical processes, but short
compared to the variation of the expansion rate. We also take modes with
physical length scales larger than the microphysical processes. The comoving
wave number will be denoted by $k$ and the physical wave number $q=ka^{-1}$. 

The Langevin equation for the fourier transform of the inflaton fluctuations
$\eta_k$ takes the form
\cite{berera00}
\begin{equation}
\ddot\eta_k+3H\dot\eta_k+\Gamma\dot\eta_k+
k^2a^{-2}\eta_k=\xi_k,\label{stoch}
\end{equation}
where $\xi_k$ is a stochastic noise source. Over the timescales of interest,
$H$ and $\Gamma$ are constant and $a=\exp(Ht)$. As in the previous section, we
set $r =\Gamma/3H$.

The correlation function of the noise can be found by converting flat space
results \cite{gleiser94} to the comoving frame. In the high temperature limit
$T\to\infty$, the noise source is Markovian,
\begin{equation}
\langle\xi_k(t)\xi_{-k'}(t')\rangle=2\Gamma T a^{-3}
(2\pi)^3\delta^3(k-k')\delta(t-t').\label{noise}
\end{equation}
The scale factor appears due to the use of comoving coordinates. (The
temperature should be large in comparison with the mass of the particles which
make up the radiation for this approximation. We shall make the same
approximation for the damping term later.)

For the eventual comparison with observational data, we define the power
spectra ${\cal P}(k)$ by
\begin{eqnarray}
{\cal P}_{\phi\phi}(k)(2\pi k^{-1})^3\delta^3(k-k')&=&
\langle\eta_k(t)\eta_{-k'}(t)\rangle\\
{\cal P}_{\pi\pi}(k)(2\pi k^{-1})^3\delta^3(k-k')&=&
\langle\dot\eta_k(t)\dot\eta_{-k'}(t)\rangle\\
{\cal P}_{\pi\phi}(k)(2\pi k^{-1})^3\delta^3(k-k')&=&
\langle\dot\eta_k(t)\eta_{-k'}(t)\rangle
\end{eqnarray}
When comparing to classical perturbation theory it is also useful to define
root-mean-square fluctuation amplitudes
\begin{eqnarray}
\delta\phi(k)&=&|{\cal P}_{\phi\phi}(k)|^{1/2}\\
\delta\dot\phi(k)&=&|{\cal P}_{\pi\pi}(k)|^{1/2}
\end{eqnarray}
Defined in this way, $\delta\phi$ has the same dimensions as $\phi$.

The power spectra can be found by solving the stochastic equation (\ref{stoch})
by green function methods,
\begin{equation}
\eta_k(t)=-{\pi\over H^2}\int_z^\infty\left\{
J_\nu(z)Y_\nu(z')-J_\nu(z')Y_\nu(z)\right\}
\left({z\over z'}\right)^\nu\xi(z'){dz'\over z'}
\end{equation}
where $z(t)=k/a(t)H$ and the order of the Bessel functions is
\begin{equation}
\nu={3\over 2}(1+r).
\end{equation}
It is now possible to use the relation (\ref{noise}) to obtain the power
spectrum, 
\begin{equation}
{\cal P}_{\phi\phi}(k)=
-\left({\pi\over 2}\right)^2 2\Gamma T 
\int_z^\infty \left\{J_\nu(z)Y_\nu(z')-J_\nu(z')Y_\nu(z)\right\}^2
z^{2\nu} z^{\prime2-2\nu}dz'.
\end{equation}
Similarly, for the time derivatives,
\begin{equation}
{\cal P}_{\pi\pi}(k)=-\left({\pi\over 2}\right)^2 H^2z^22\Gamma T 
\int_z^\infty \left\{J_{\nu-1}(z)Y_\nu(z')-J_\nu(z')Y_{\nu-1}(z)\right\}^2
z^{2\nu} z^{\prime2-2\nu}dz'.
\end{equation}

The formula for the power spectrum can be simplified at late or early times.
For early times, when the physical size of the mode $a/k$ is much smaller than
the horizon, we can use the $z\gg \nu$ approximation for the Bessel functions
to obtain
\begin{eqnarray}
{\cal P}_{\phi\phi}(k)&\sim&ka^{-1}T\\
{\cal P}_{\pi\pi}(k)&\sim&k^3a^{-3}T
\end{eqnarray}
Allowing for the transformation from comoving to physical wavenumbers, these
results are identical to the thermal fluctuations of a field in flat space. The
rms fluctuations are proportional to $T^{1/2}$, which is typical of the
fluctuations of a classical object in a radiation field \cite{lifshitz}.

A different approximation holds for later times, when $H<ka^{-1}<(\Gamma
H)^{1/2}$, and we can use the $z\ll \nu$ approximation for the Bessel
functions. The power spectrum simplifies to
\begin{equation}
{\cal P}_{\phi\phi}(k)\sim{\pi^2\over 2}\Gamma Tz^{2\nu}Y_\nu(z)^2
\int_0^\infty J_\nu(z')^2
z^{\prime2-2\nu}+O(z^3)
\end{equation}
The root-mean-square fluctuation is approaching the dominant solution 
$z^{\nu}Y_\nu(z)$ of the homogeneous equation for $\eta_k$ in this regime as if
the noise had no effect. With the warm inflation condition $\Gamma\gg H$, the
results further reduce to
\begin{eqnarray}
{\cal P}_{\phi\phi}(k)&\sim&\left({\pi\over 4}\right)^{1/2}(\Gamma H)^{1/2}T
\left(1+{1\over \Gamma H}{k^2\over a^2}+\dots\right)\\
{\cal P}_{\pi\pi}(k)&\sim&\left({\pi\over 4}\right)^{1/2}{k^4\over \Gamma^2
a^4}(\Gamma H)^{1/2}T
\end{eqnarray}
The cross-correlation reduces to
\begin{equation}
{\cal P}_{\pi\phi}(k)\sim -({\cal P}_{\phi\phi}(k))^{1/2}
({\cal P}_{\pi\pi}(k))^{1/2},\label{cross}
\end{equation}
which also indicates that the effects of the noise term are fading away. 

The transition between the two regimes has been called freezeout
\cite{berera00},
and occurs at a particular time $t_F(k)$ when $ka^{-1}=(\Gamma H)^{1/2}$. In
warm inflation, the freezeout time will always preceed the hubble crossing
time, at which $ka^{-1}=H$.  The evolution of the inflaton becomes
increasingly deterministic for times $t>t_F$.

The fluctuations approach a constant value,
\begin{eqnarray}
\delta\phi(k)&\sim&\left(\pi\over 4\right)^{1/4}
(\Gamma H)^{1/4}T^{1/2}\label{dphi}\\
\delta\dot\phi(k)&\sim&-\left({\pi\over 4}\right)^{1/4}
{k^2\over \Gamma a^2}(\Gamma H)^{1/4}T^{1/2}.\label{ddphi}
\end{eqnarray}
Apart from the numerical factor, the value of $\delta\phi$ agrees with the
results of Berera \cite{berera00}, who analysed equation (\ref{stoch}) in the
regime $ka^{-1}>(\Gamma H)^{1/2}$. The sign of $\delta\dot\phi$ has been chosen
for consistency with the cross correlations (\ref{cross}). It is also
consistent with the deterministic solution.

\subsection{Energy fluctuations}

Now we turn to thermal fluctuations in the energy density in flat space. The
classical theory of energy fluctuations is based on fundamental principles
of statistical physics. For the energy fluctuation $\Delta E$ in a fixed
volume $V$, Lifshitz and Pitaevstii \cite{lifshitz} give the formula
\begin{equation}
\langle(\Delta E)^2\rangle_V=C_v T^2,
\end{equation}
where $C_v$ is the specific heat. This result can also be expressed in terms of
the energy density $\rho$ and the entropy density $s$,
\begin{equation}
\langle(\Delta\rho)^2\rangle_V= V^{-1}T^3s_{,T}.\label{thermo}
\end{equation}
We shall investigate the equivalent result in finite temperature quantum field
theory before considering the application to warm inflation.

Consider a scalar field $\chi$ in thermal equilibrium at temperature $T$.
Fluctuations $\varepsilon$ in the energy density are defined by
\begin{equation}
\varepsilon={\cal H}(x)-\langle{\cal H}\rangle_T,
\end{equation}
where the energy density operator
\begin{equation}
{\cal H}=\frac12\dot\chi^2+\frac12(\nabla\chi)^2+\frac12 m^2\chi^2,
\end{equation}
and the energy density $\rho_r=\langle{\cal H}\rangle_T$ is the thermal
ensemble expectation value of ${\cal H}$.

The correlation function
\begin{equation}
\langle\varepsilon(x)\varepsilon(x')\rangle_T
=\langle{\cal H}(x){\cal H}(x')\rangle_T-
\langle{\cal H}\rangle_T^2.
\end{equation}
The simplest way to remove any divergences will be to subtract the zero
temperature correllations, which in any case are not of interest to us. We
therefore define the power spectrum of the radiation ${\cal P}_r$ by
\begin{equation}
{\cal P}_r(q)(2\pi q^{-1})^3\delta^3(q-q')
=\langle\varepsilon_{q}(t)\varepsilon_{-q'}(t')\rangle_T
-\langle\varepsilon_{q}(t)\varepsilon_{-q'}(t')\rangle_0.
\end{equation}
As before, we also define a root-mean-square fluctuation
\begin{equation}
\delta\rho_r(k)=|{\cal P}_r(q)|^{1/2}.
\end{equation}

The expectation values for four field operators can be reduced using the
cluster decomposition principle \cite{weinberg} to give 
\begin{equation}
{\cal P}_r(q)
=\int{d^3q'\over (2\pi)^3}\,q^3(-q(q-q')+\partial_t\partial_{t'}+m^2)^2
(G_T^>(q,t)G_T^>(q-q',t')-G_0^>(q,t)G_0^>(q-q',t')),
\end{equation}
where the expectation values of products of two fields is given by the thermal
green function,
\begin{equation}
\langle\chi_q(t)\chi_{-q'}(t')\rangle=(2\pi)^2\delta^3(q-q')G_T^>(q,t-t').
\end{equation}
For a bosonic field in thermal equilibrium,
\begin{equation}
G_T^>(q,t)=n(\omega_q)e^{i\omega_qt}+(1+n(\omega_q))e^{-i\omega_qt},
\end{equation}
where $\omega_q=(m^2+q^2)^{1/2}$ and $n(\omega)=(e^{\omega/T}-1)^{-1}$.
The power spectrum reduces to the integral
\begin{equation}
{\cal P}_r(q)
=\int{d^3q'\over (2\pi)^3}\,{q^3\over 4\omega_{q'}\omega_{q-q'}}
\left(2\omega_{q-q'}^2\omega_{q'}^2-m^2q^2\right)
\left(2n(\omega_{q-q'})n(\omega_{q'})+n(\omega_{q-q'})+n(\omega_{q'})\right).
\end{equation}

The classical limit corresponds to $q\ll T$. The integral is then dominated by
the region $q'\gg q$. For a massless field, the integral reduces to
\begin{equation}
{\cal P}_r(q)=4q^3T\rho_r\label{pr},
\end{equation}
where $\rho_r$ is the energy density. This is in agreement with the
thermodynamic result (\ref{thermo}) for pure radiation if we identify $q^3$
with $V^{-1}$.

In the expanding universe, the inflaton and radiation fluctuations are
effectively in equilibrium for times $t<t_F$, where $t_F$ is the freezeout
time mentioned earlier. The total energy fluctuations in this regime are given
by equation (\ref{thermo}). After $t_F$, the radiation and inflaton
fluctuations are uncorrelated and equation (\ref{pr}) can be used for the
fluctuations in the energy density of the radiation,
\begin{equation}
\delta\rho_r(k)=\left({2\pi^2\over
15}\right)^{1/2}g_*^{1/2}k^{3/2}a^{-3/2}T^{5/2},
\label{drho}
\end{equation}
where $g^*$ is the effective particle number and we have introduced the
comoving wave number $k=aq$. 

The total energy is defined by equation (\ref{density}). The inflaton energy
fluctuations for $t>t_F$ are therefore
\begin{equation}
\delta\rho_\phi=\dot\phi\,\delta\dot\phi+V_{,\phi}\delta\phi,
\end{equation}
where $\delta\phi$ and $\delta\dot\phi$ are given above.

\section{Cosmological Perturbations}

The thermal fluctuations occuring during warm inflation evolve gradually into
cosmological perturbations. The thermal noise has a diminishing effect on the
development of the perturbations until the freezeout time when it becomes
insignificant. The perturbations can then be described by a gaussian random
field which grows deterministically.  

In warm inflation there are cosmological perturbations in the inflaton
field, the radiation and the gravitational field. We shall only consider the
scalar gravitational mode, which means that there is one degree of freedom in
the metric perturbations. We shall also use a particular gauge, the zero-shear
gauge, in which the scalar metric perturbation has the form
\begin{equation}
ds^2=-(1-2\varphi)dt^2+a^2(1+2\varphi)\delta_{ij}dx^idx^j.
\end{equation}
The inflaton perturbation will be denoted by $\delta\phi$, the temperature
perturbation $\delta T$ and the velocity perturbation $v$.

For any random perturbation field $g(x)$, the power spectrum is defined by an
average 
\begin{equation}
{\cal P}_g(k)(2\pi k^{-1})^3\delta^3(k-k')=
\langle g_kg_{-k'}\rangle,
\end{equation}
where $g_k$ is the fourrier transform of $g(x)$. The amplitude will be
normalised by
\begin{equation}
g(k)=|{\cal P}_g(k)|^{1/2}.
\end{equation}
Note that $g(k)$ satisfies the same linear equations of motion as $g_k$. We use
this normalisation for the perturbations and omit explicit reference to $k$.

The total energy density and pressure perturbations are then
\begin{eqnarray}
\delta\rho&=&\dot\phi\,\delta\dot\phi+V_{,\phi}\delta\phi
+\dot\phi^2\varphi+T\delta s\label{drhot}\\
\delta p&=&\dot\phi\,\delta\dot\phi-V_{,\phi}\delta\phi
+\dot\phi^2\varphi+s\delta T
\end{eqnarray}
and the energy momentum flux $(\rho+p)v$.

A complete set of perturbation equations can be obtained from the Einstein
field equations and the scalar field equation \cite{oliveira01,hwang02}. From
the Einstein equation we obtain
\begin{eqnarray}
\dot\varphi+H\varphi+4\pi Gk^{-1}a(p+\rho)v&=&0\label{onep}\\
3H\dot\varphi+(3H^2+k^2a^{-2})\varphi-4\pi G\delta\rho&=&0\label{twop}\\
\ddot\varphi+4H\dot\varphi+(2\dot H+3H^2)\varphi+
4\pi G\delta p&=&0.\label{threep}
\end{eqnarray}
Perturbations of the scalar field equation give
\begin{equation}
\delta\ddot\phi+(3H+\Gamma)\delta\dot\phi+\dot\phi(\delta\Gamma)
+k^2a^{-2}\delta\phi+\delta V_{,\phi}+
4\dot\phi\dot\varphi-\Gamma\dot\phi\varphi-2V_{,\phi}\varphi=0.\label{fourp}
\end{equation}
In general, we can have $\Gamma\equiv\Gamma(\phi,T)$ and $V\equiv V(\phi,T)$,
\begin{eqnarray}
\delta\Gamma&=&\Gamma_{,\phi}\delta\phi+\Gamma_{,T}\delta T\\
\delta V_{,\phi}&=&V_{,\phi\phi}\delta\phi+V_{,\phi T}\delta T\\
\delta s&=&-V_{,\phi T}\delta\phi-V_{,TT}\delta T.
\end{eqnarray}
The numerical results of the next section are obtained
by integrating these equations with the initial conditions set by the thermal
fluctuations.

During warm inflation, the background solution rapidly approaches the
slow-rolldown approximation (\ref{slowp}-\ref{slowh}). In a similar way, the
perturbations also have a slow-roll limit, which begins when their length
scales become larger than the horizon, $k<aH$. The $\delta\ddot\phi$,
$\dot\varphi$ and $\ddot\varphi$ terms become insignificant, and perturbations
approach a slow-roll large-scale limit
\begin{eqnarray}
\delta\phi&\sim&-C H^{-1}\dot\phi\label{slowdp}\\
\varphi&\sim&4\pi GC(1+r )\dot\phi^2\label{slowg}\\
v&\sim& -C ka^{-1}H^{-1}\label{slowv}
\end{eqnarray}
where $C$ is a constant and $r =\Gamma/3H$.

The cosmological perturbations can also be described in terms of gauge
invariant quantities, and these are useful for following the development of
perturbations after the end of inflation. The curvature perturbation ${\cal R}$
\cite{lukash80}, is defined by
\begin{equation}
{\cal R}=\varphi-k^{-1}aHv\label{defphi},
\end{equation}
The entropy perturbation $e$ is defined by \cite{kodama84}
\begin{equation}
e=\delta p-c_s^2\delta\rho,
\end{equation}
where $c_s^2=\dot p/\dot \rho$. In the slow-roll large-scale limit, the
curvature perturbation is
constant, ${\cal R}\sim C$ and the entropy perturbations vanish.

The curvature and entropy perturbations are
coupled by a second order equation
\cite{hwang02}
\begin{equation}
\ddot{\cal R}+\Gamma_{\cal R}\dot{\cal R}
+c_s^2k^2a^{-2}{\cal R}=-((He_1)^\cdot-\Gamma_{\cal
R}He_1),\label{gaugeinv}
\end{equation}
where $e_1=e/(p+\rho)$ and
\begin{equation}
\Gamma_{\cal R}=
-3Hc_s^2-2H^{-1}\dot H-2c_s^{-1}\dot c_s.
\end{equation}
The perturbation equation (\ref{gaugeinv}) implies the classic result
\cite{bardeen83}, that if
there is no further entropy generation then the curvature fluctuation ${\cal
R}$ will retain the constant value it reached during inflation whilst
the wavelength exceeds the horizon size, $k<aH$. 
  
If the entropy perturbation vanishes on small as well as large scales, then we
only need to solve the cosmological perturbation equation for ${\cal R}$. This
is a reasonable approximation during supercooled inflation, and the equation is
often
used in that context \cite{leach01}.

In warm inflation the entropy perturbations are important. Nevertheless, for
the
case $\Gamma_{,T}=0$, it is possible to obtain analytic results. It is
advantageous to use the inflaton equation for approximate solutions, relying
on the fact that the metric perturbations remain small on scales smaller than
the horizon, $k>aH$. An analytic approximation to the density perturbation
amplitude can be obtained by matching the classical result for $\delta\phi$
Eq. (\ref{slowp}) to the thermal fluctuations Eq. (\ref{dphi}) at the crossing
time
$t_H$ when $k=aH$,
\begin{equation}
{\cal P}_{\cal R}\sim\left({\pi\over 4}\right)^{1/2}
{H^{5/2}\Gamma^{1/2}T\over\dot\phi^2}.\label{anr}
\end{equation}
This result is analogous to the result ${\cal P}_{\cal R}=H^4/\dot\phi^2$ for
supercooled inflation.

The spectral index $n_s$ is defined by
\begin{equation}
n_s-1={\partial \ln{\cal P}_{\cal R}\over\partial \ln k}.
\end{equation}
The slow-roll equations (\ref{hdot}-\ref{tdot}) enable us to express the
spectral index for the amplitude (\ref{anr}) in terms of slow-roll
parameters,
\begin{equation}
n_s-1={1\over r }\left(-{9\over 4}\epsilon+{3\over 2}\eta-
{9\over 4}\beta\right).\label{speci}
\end{equation}
The first two terms agree with reference \cite{berera00}. The $\beta$ term
shows the dependence of the spectrum on the gradient of the reheating term.
For comparison, the spectral index for standard, or supercooled inflation, is
$n_s=1-6\epsilon+2\eta$ \cite{liddle92}.

\section{Numerical Results}

We shall consider the evolution of the cosmological perturbations and the
spectrum of density perturbations in a particular toy model which allows us to
investigate a number of interesting phenomena. The toy models include some
basic features of consistant models of warm inflation, such as
\cite{berera99,berera03}, but they are simplified in order to isolate
particular effects.

The same potential for the inflaton field will be used throughout. The thermal
corrections to the potential are taken to be insignificant. Three different
types of damping term will be considered. We hope to compare the perturbation
spectra for different potentials and potentials with large thermal corrections
in later work.

The potential we choose is
\begin{equation}
V(\phi)={1\over 4}\lambda (\phi^2-\phi_0^2)^2.
\end{equation}
The important features of this potential are the maximum at $\phi=0$ and the
minimum at $\phi=\phi_0$. These features can arise in models with spontaneous
symmetry breaking and have been used before in models of supercooled inflation
\cite{lyth99}.

For the damping term $\Gamma$, we will take
\begin{equation}
\Gamma=\Gamma_0\left({\phi\over \phi_0}\right)^b
 \left({T\over V(0)^{1/4}}\right)^c,\label{friction}
\end{equation}
with different choices of $b$ and $c$. In the first example, $b=c=0$, which
allows comparison with the analytic results of Taylor and Berera
\cite{taylor00}. The second example uses $b=2$ and $c=0$ to isolate the
effects of the $\phi$ dependence. The third case, $b=2$, $c=-1$ is the damping
term first calculated by Hosoya \cite{hosoya84}, and used in some consistant
models of warm inflation, for example \cite{berera99}

\subsection{Homogeneous solution}

\begin{figure}
\scalebox{0.7}{\includegraphics{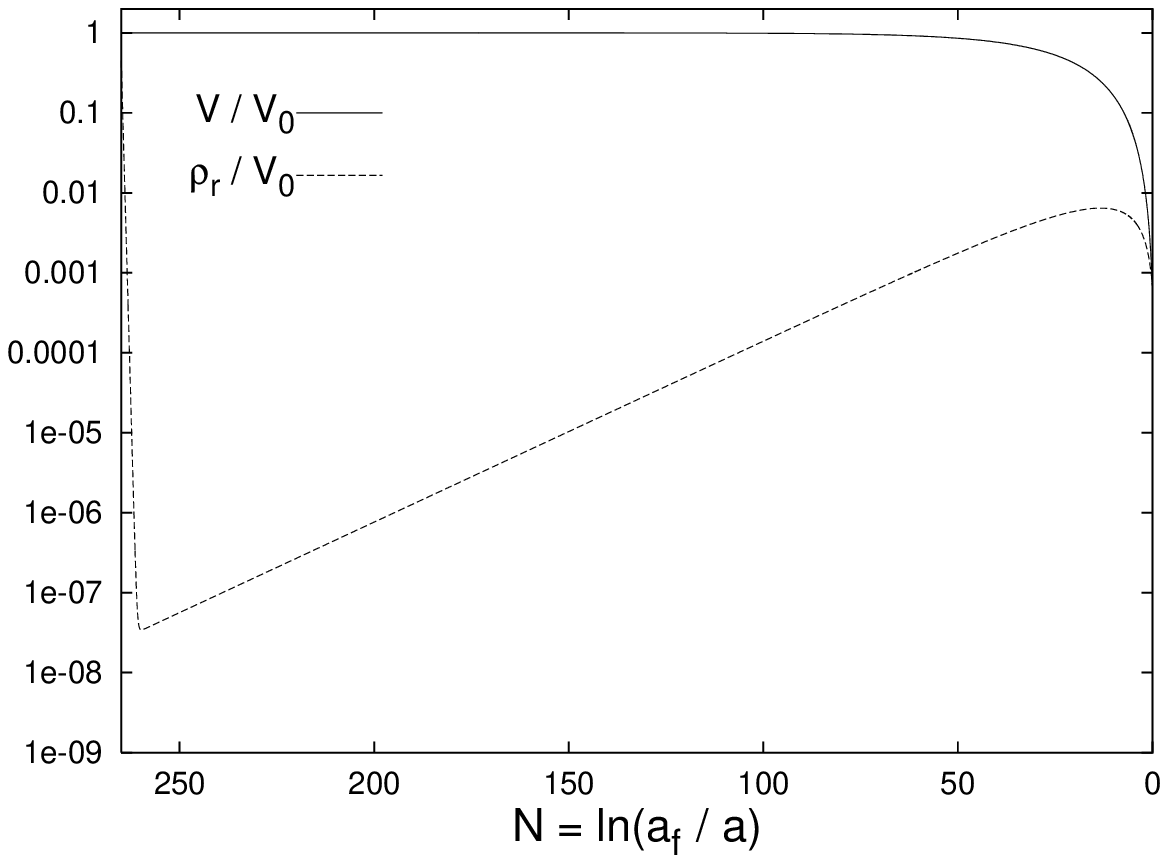}}
\scalebox{0.7}{\includegraphics{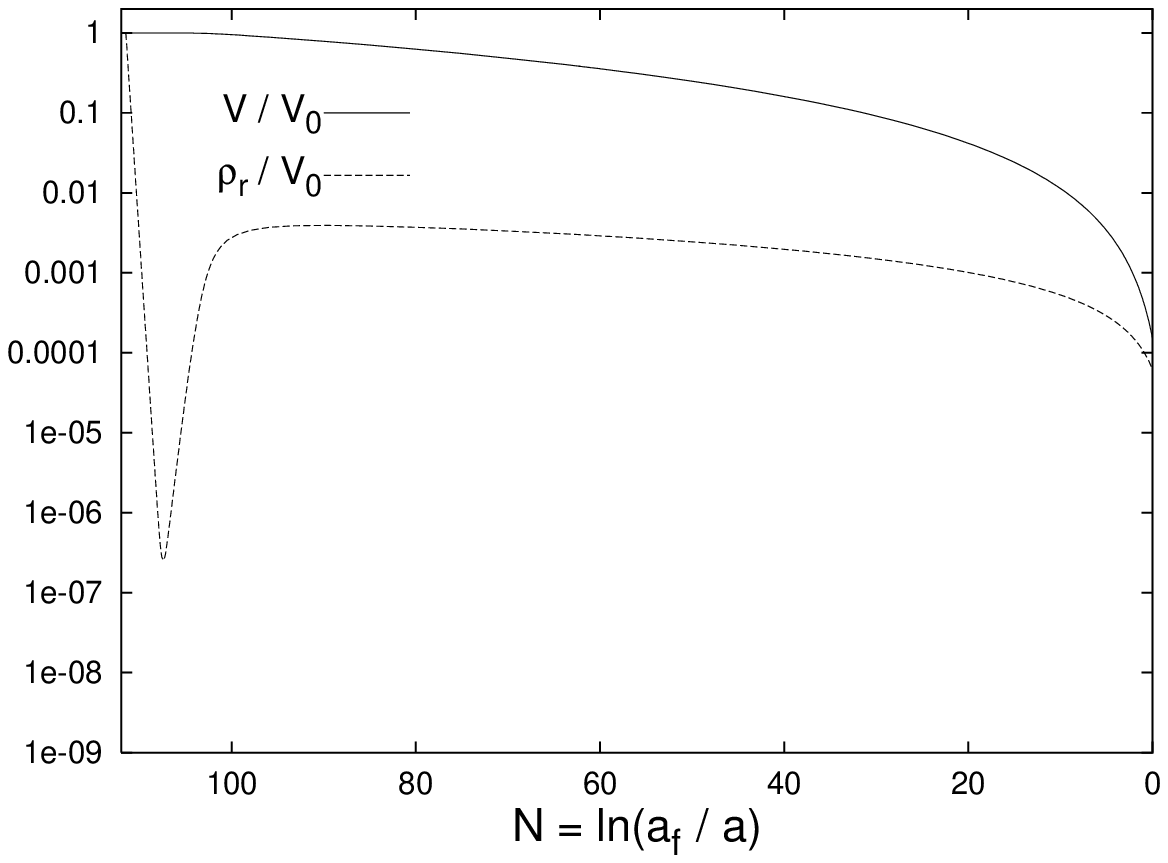}}
\scalebox{0.7}{\includegraphics{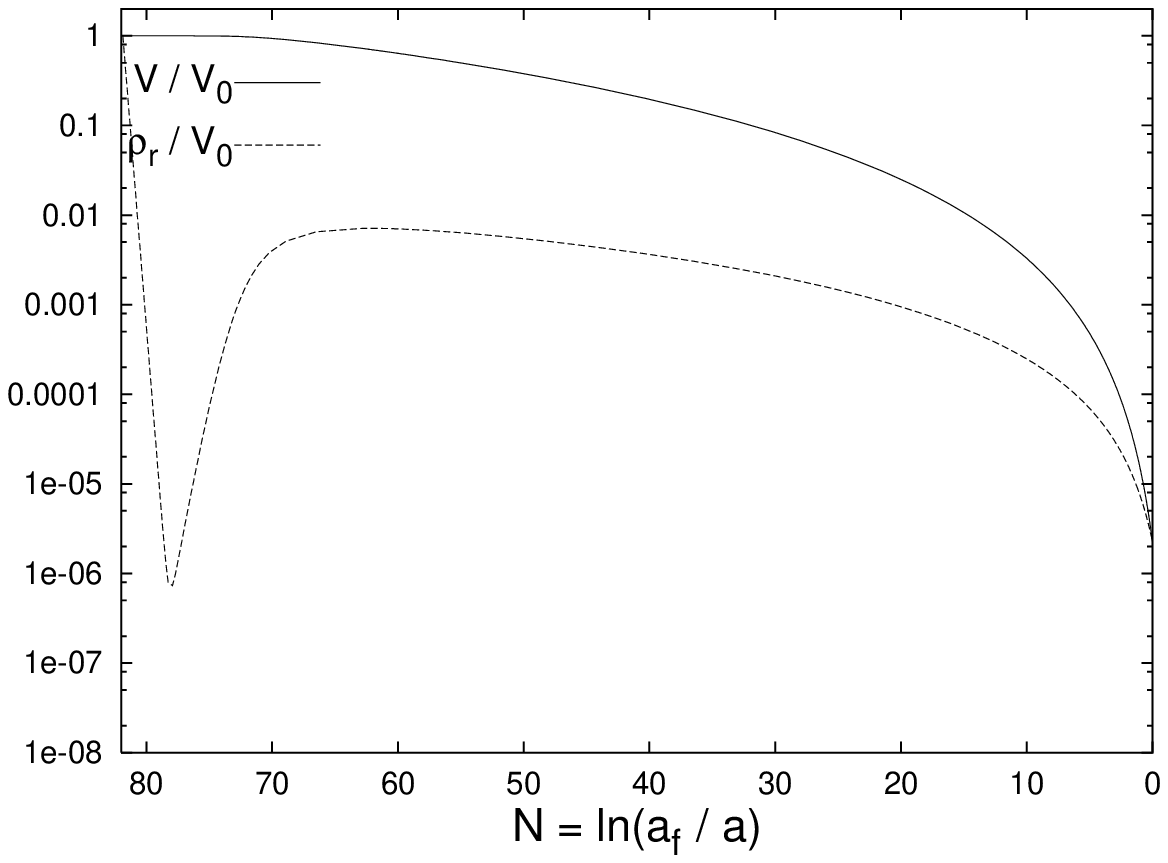}}
\caption{\label{fig1}The potential and the radiation density are plotted
against $N$ for the three models described in the text. 
(I) Constant damping, (II) Damping depends on $\phi$, (III) Damping depends on
$\phi$ and $T$. In all cases $r\approx 100$ at $N=60$.}
\end{figure}

The inflaton $\phi$, entropy $s$ and scale factor $a$ satisfy equations
(\ref{wip}-\ref{wia}). The numerical solutions shown in figure \ref{fig1} plot
the potential and the radiation density $\rho_r$ for the three models listed in
table \ref{table1}. The radiation density drops rapidly at first due to the
expansion, but rises as the solutions approach the slow-roll regime.

\begin{table}
\caption{\label{table1}Three models with different damping terms. $T_f$ is the
temperature at the end of inflation and $g_*=100$.}
\begin{ruledtabular}
\begin{tabular}{llllllll}
model&$b$&$c$&$\lambda$&$\phi_0$&$V(0)$&$\Gamma_0$&$T_f$\\
I&$0$&$0$&$2\times 10^{-15}$&$0.3m_{pl}$&
$(4.48\times10^{14}GeV)^4$&$1.43\times10^{13}GeV$&$3.0\times10^{13}GeV$\\
II&$2$&$0$&$6.4\times 10^{-16}$&$0.6m_{pl}$&
$(6.75\times10^{14}GeV)^4$&$4.3\times10^{13}GeV$&$2.1\times10^{13}GeV$\\
III&$2$&$-1$&$6.4\times 10^{-16}$&$0.6m_{pl}$&
$(6.75\times10^{14}GeV)^4$&$4.3\times10^{11}GeV$&$1.1\times10^{13}GeV$\\
\end{tabular}
\end{ruledtabular}
\end{table}

The end of inflation happens when the slow-roll parameter $\epsilon=1+r$,
and coincides with the equality between radiation and potential energy,
$\rho_r=V$. It is convenient to relate the times of events during inflation to
the time of the end of inflation, $t_f$, and so we plot $N(t)$, where
\begin{equation}
N=\ln(a_f/a),
\end{equation}
along the horizontal axis. 

The radiation temperatures at the end of inflation for the three models are
given in table \ref{table1}. These determine the Hubble length at the end of
inflation (when the total energy density is $2\rho_r$),
\begin{equation}
cH_f^{-1}=1.77\times 10^{-25} g_*^{-1/2}T_{14}^{-2}\,\hbox{m}
\end{equation}
where $T_{14}=T_f/(10^{14}GeV)$. For comparison, a comoving scale which is 500
Mpc at the present epoch would have length
\begin{equation}
3.59\times 10^{-2}T_{14}^{-1}\,\hbox{m}
\end{equation} 
at the end of inflation.

In the models with $b=2$, the inflation continues for larger values of $\phi$
than the model with $b=0$. A consequence of this is that the reheat
temperature is smaller in these models. This can have an important effect on
the production of gravitinos \cite{liddle}.

\subsection{Perturbations on small scales}

Our earlier discussion of thermal fluctuations ignored the effects of
fluctuations in the metric and the damping term. The cosmological perturbation
equations imply that the metric perturbation remains small and has little
effect before the horizon crossing time $t_H$, when $k>aH$. However, for an
accurate treatment, and for dealing with fluctuations in the damping term, we
must solve the perturbed Einstein equations and perturbed equation of motion.

Our numerical code simultaneously integrates the equations for the homogeneous
background and the perturbation equations for a given wave number $k$.
The integration begins with the homogeneous solution alone. Integration of the
perturbation equations only commences at the freezeout time $t_F$, defined by
the condition $k=a(\Gamma H)^{1/2}$. This is done in order to ensure that the
noise term in the Langevin equation which we saw earlier can be neglected.
When the integration of the perturbation equations begins, the wavelength of
the perturbation is much smaller than the horizon size, and horizon crossing
occurs at a later time $t_H$. The integration is stopped when the radiation
energy density reaches the same value as the vacuum energy density.

Initial conditions have to be set for the perturbed quantities at the freezeout
time when the integration of the perturbation equations begins. The initial
inflaton fluctuations were given in Eq. (\ref{dphi}) and Eq. (\ref{ddphi}). The
initial value of the metric perturbation is given by the perturbed Einstein
equations (\ref{onep}) and (\ref{twop}),
\begin{equation}
\varphi={a^2\over k^2}4\pi G\,\delta\rho,
\end{equation}
where $\delta\rho$ is given by equation (\ref{drho}). Interestingly, at the
freezout time, $|\delta\rho_r|\approx|\delta\rho_\phi|\approx|(p+\rho)v|$.

\begin{figure}
\scalebox{0.7}{\includegraphics{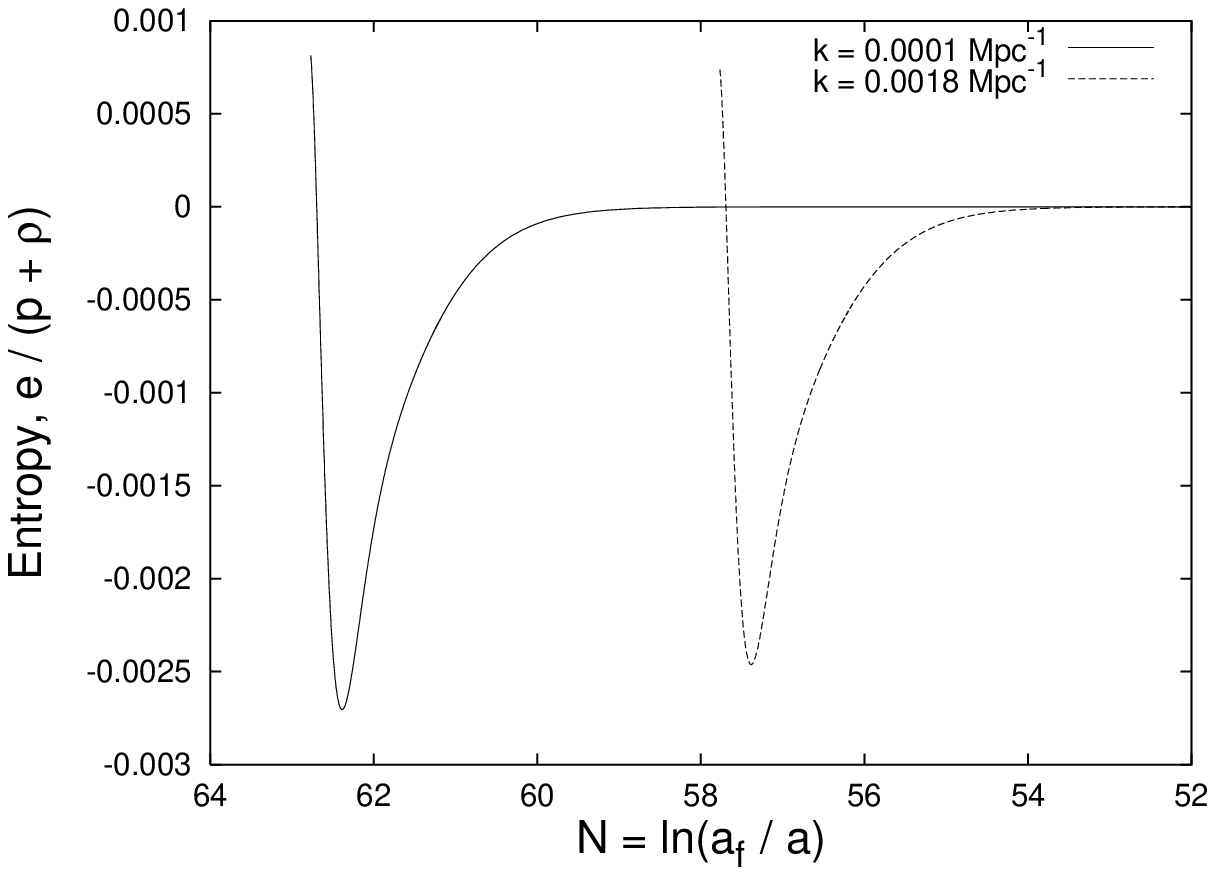}}
\scalebox{0.7}{\includegraphics{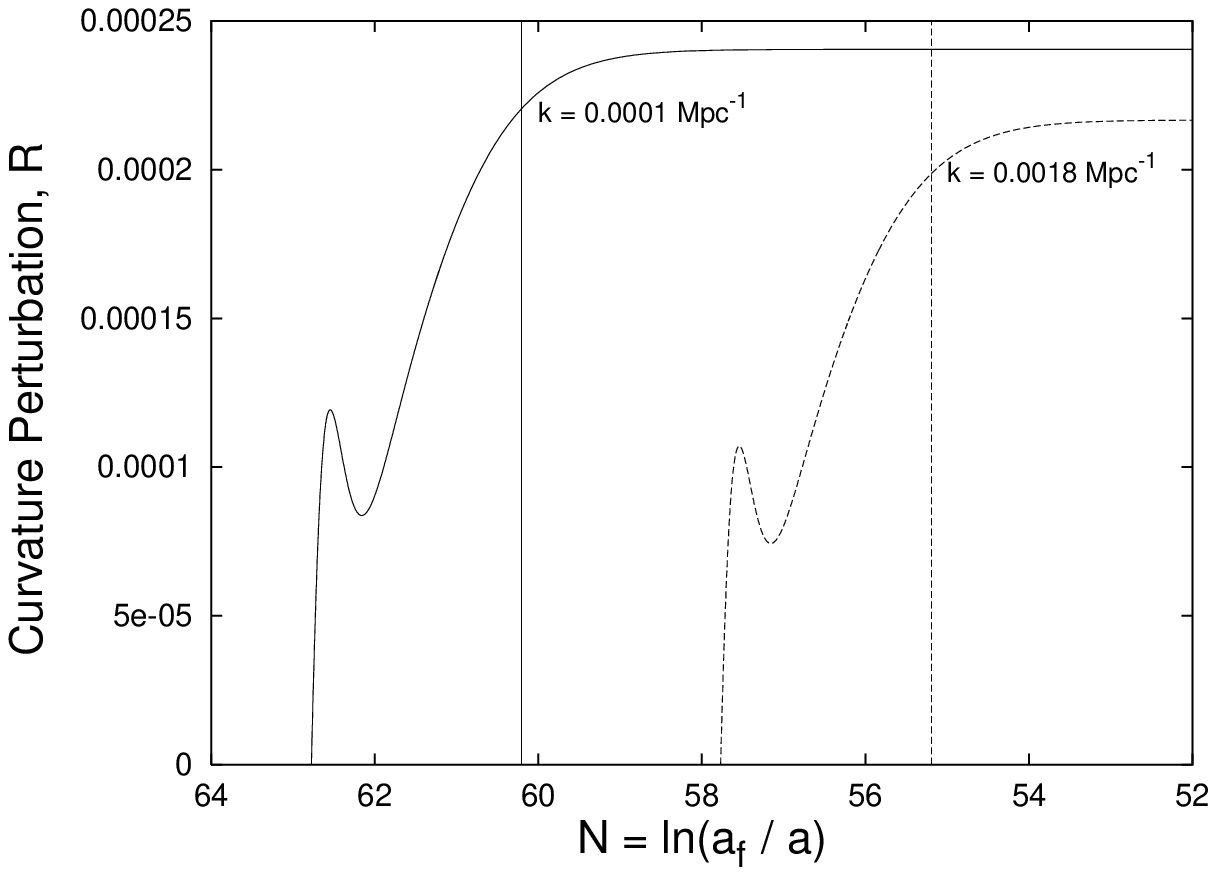}}
\scalebox{0.7}{\includegraphics{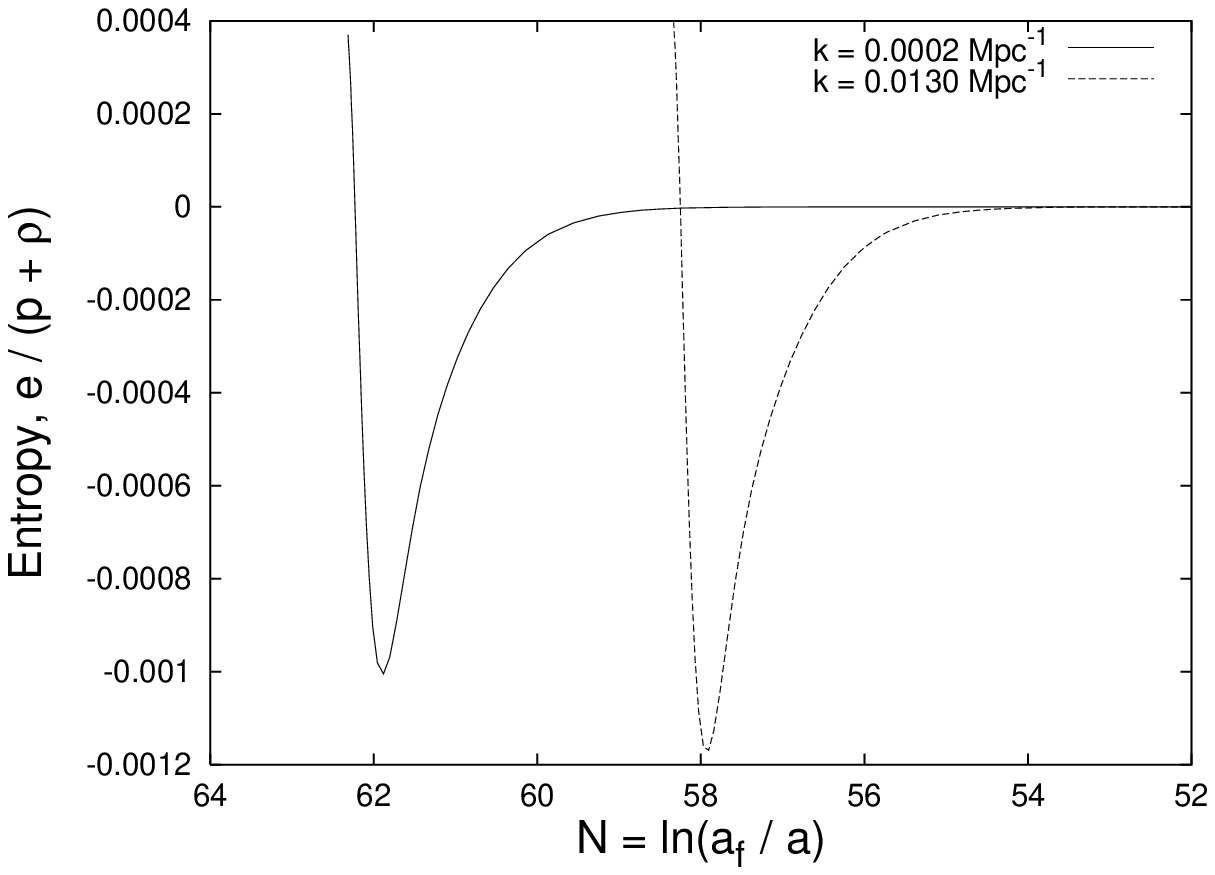}}
\scalebox{0.7}{\includegraphics{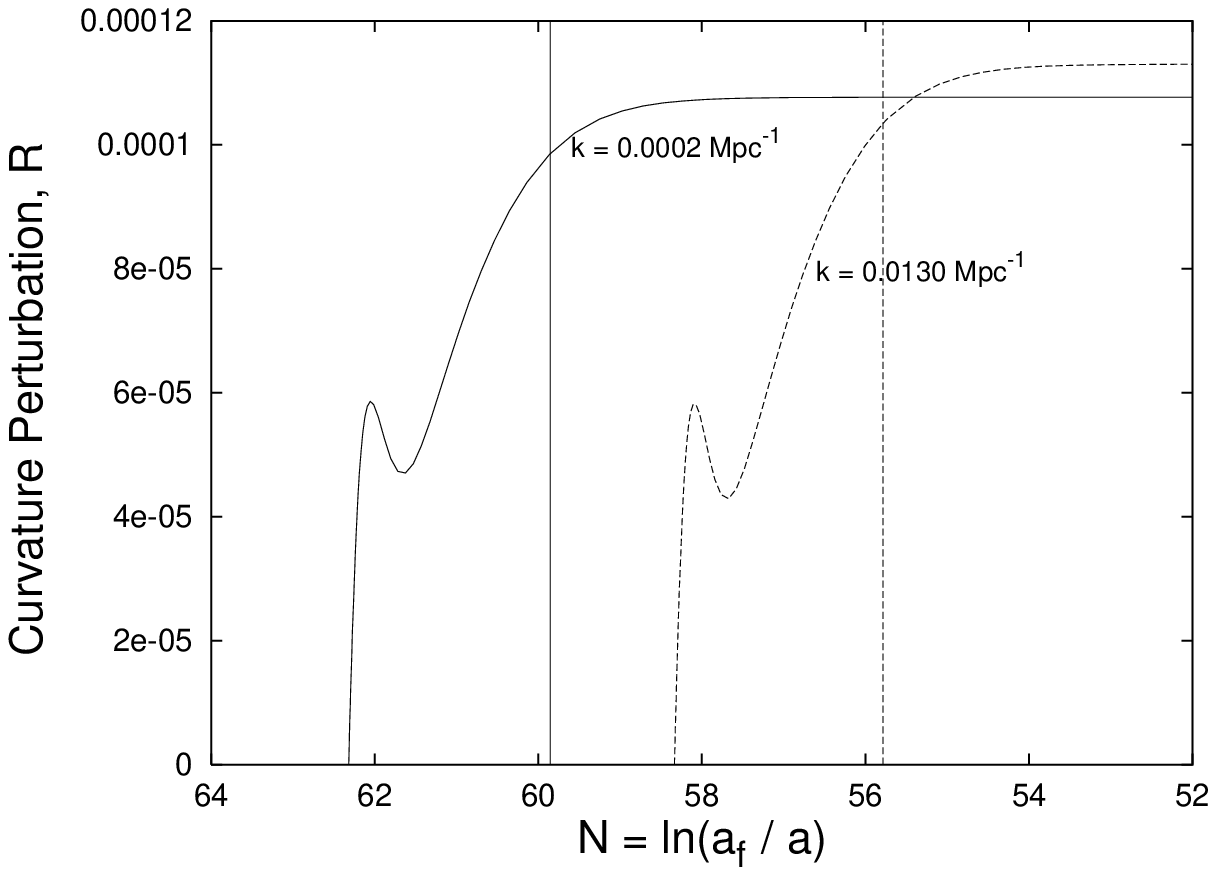}}
\scalebox{0.7}{\includegraphics{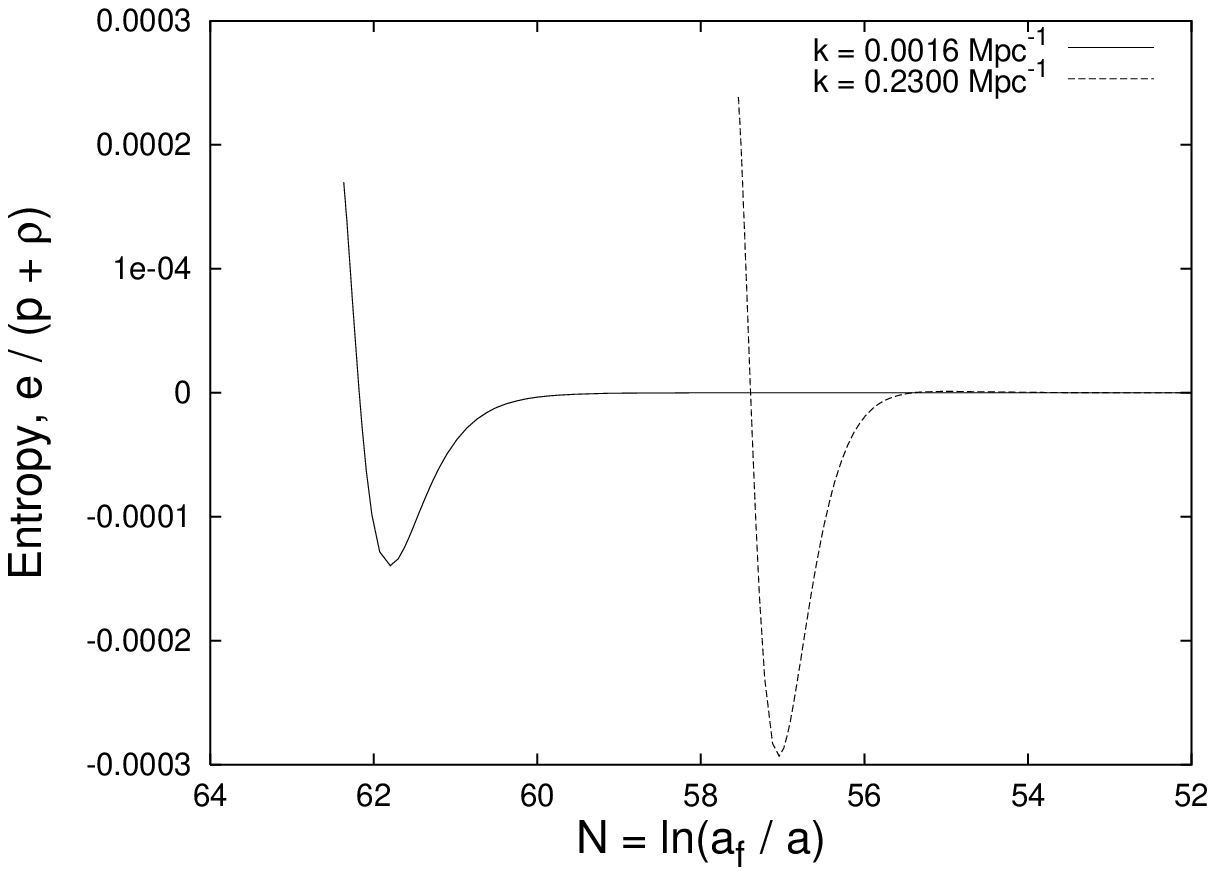}}
\scalebox{0.7}{\includegraphics{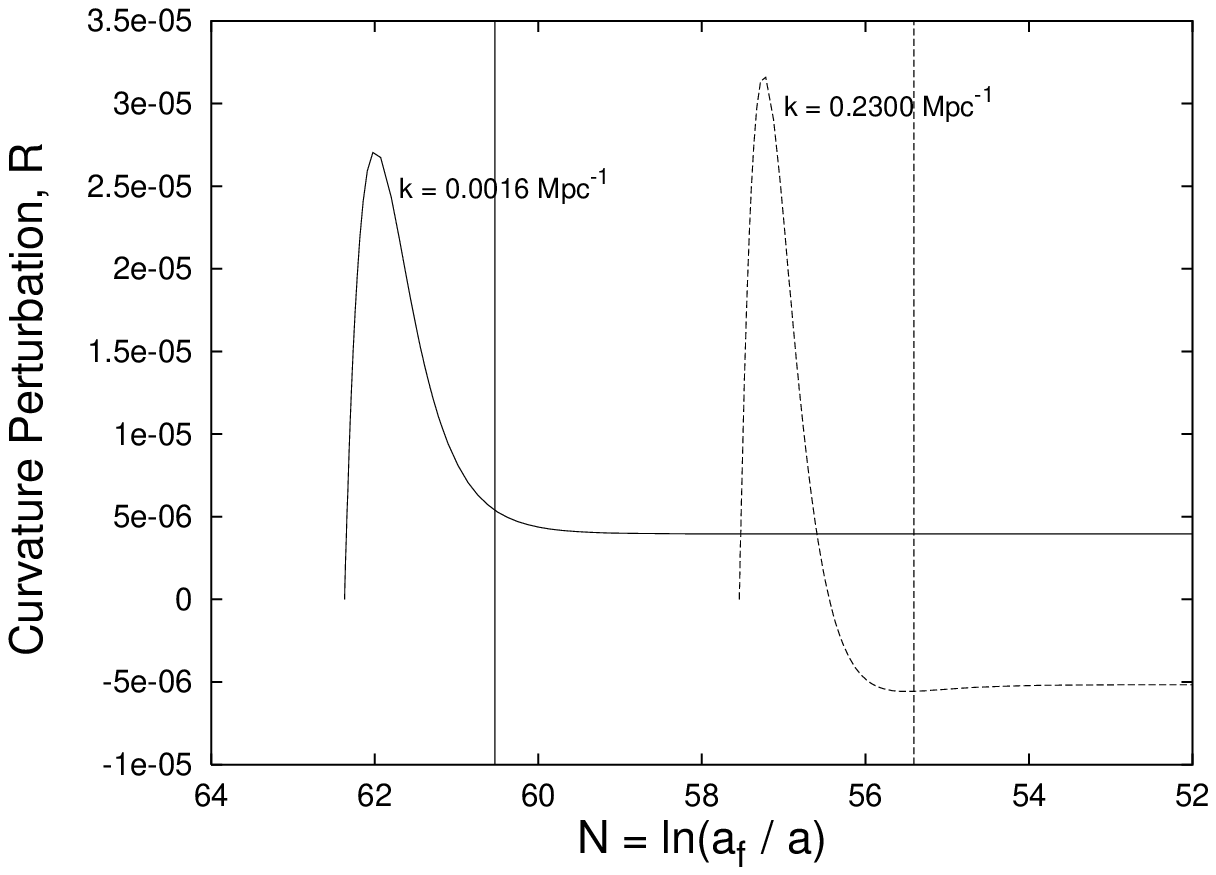}}
\caption{\label{fig2}Evolution of the fluctuations for two different
values of the comoving wavenumber $k$. The three rows represent
the three models, constant damping (top), damping depends on $\phi$ (middle),
damping depends on
$\phi$ and $T$ (bottom). The vertical lines indicates horizon crossing,
$k=aH$.}
\end{figure}

The time evolution of the gauge invariant parameters describing the
perturbations is shown in figure \ref{fig2}. The
entropy perturbations are on the left. In the range $t_F<t<t_f$, the entropy
perturbation is dominated by the radiation,
\begin{equation}
e\approx \delta (Ts).
\end{equation}
The entropy perturbation has a significant peak but always approaches zero
after the perturbation crosses the horizon. In supercooled inflation the
entropy perturbation is small compared to the curvature perturbation.

Figure \ref{fig2} also shows the amplitude of the curvature fluctuations for
the three damping models. The amplitudes begin relatively small and grow to
approach a constant value very shortly after horizon crossing. The influence of
the entropy perturbations can be seen in a small bump on the rising edge of
the plots. The final value of the curvature fluctuation in model I is within
$5\%$ of the analytic approximation (\ref{anr}).

The third model, where the damping term depend on the temperature, differs from
the other two. In this case, the entropy fluctuation has a direct influence on
the inflaton fluctuation. This causes the curvature perturbation to follow the
entropy perturbation more closely, and and even lead to the final value of the
curvature perturbation changing sign. (Note that the root mean square value of
the the curvature perturbation is given by the modulus in this case).    
The analytic approximation for the scalar perturbation amplitude is not
applicable to this case and the numerical results confirm this.

\subsection{Perturbation spectra}

\begin{figure}
\scalebox{0.7}{\includegraphics{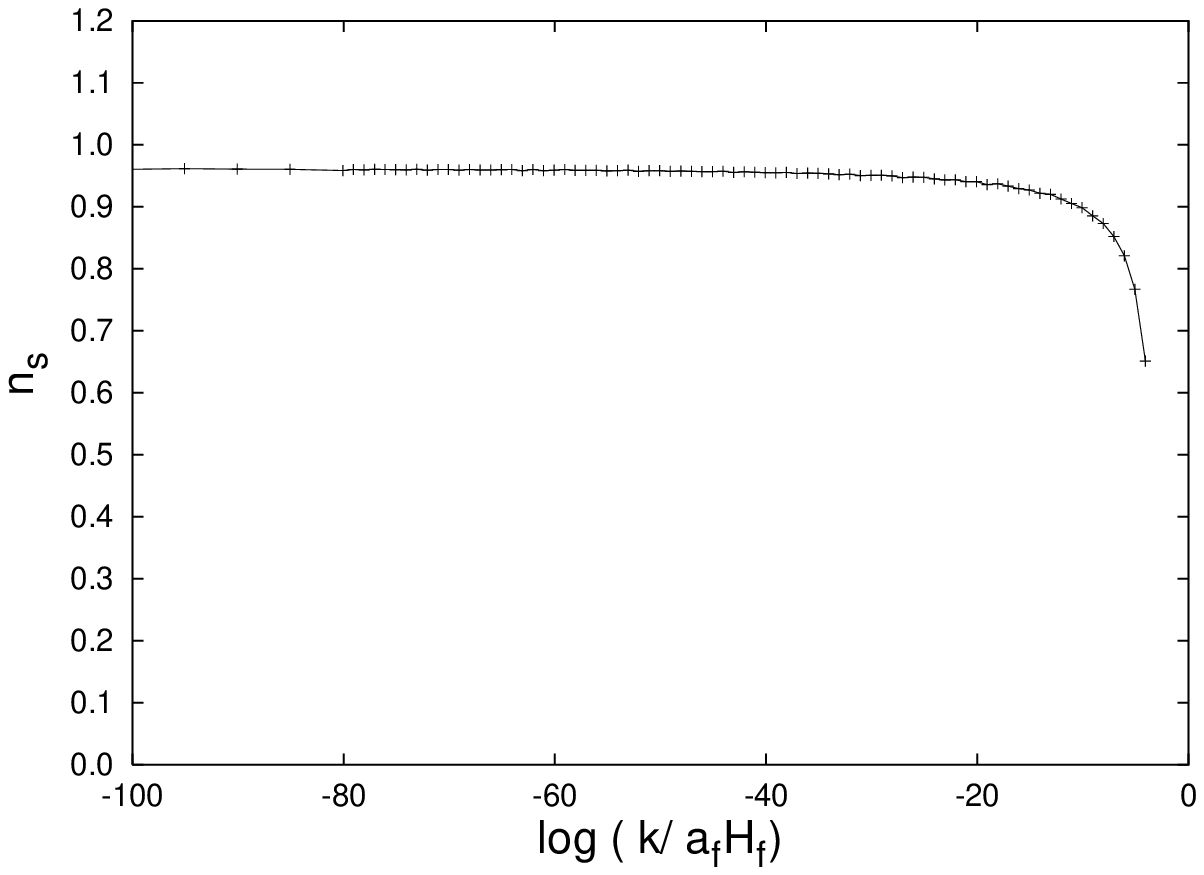}}
\scalebox{0.7}{\includegraphics{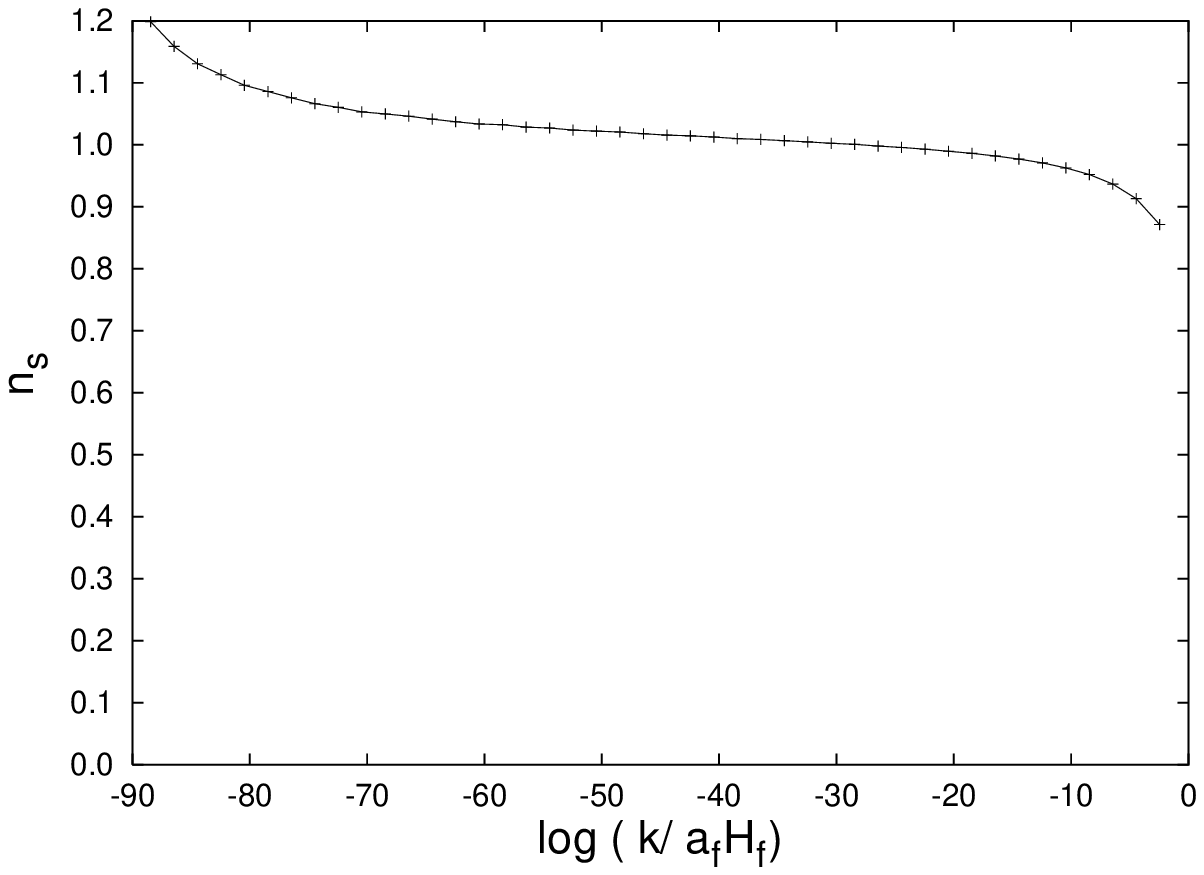}}
\caption{\label{fig4}The spectral index is plotted for the potential
$\frac14\lambda(\phi^2-\phi_0^2)^2$ with 
(I) Constant damping, (II) Damping depends on $\phi$. The physical wavenumber
$k/a$ has been normalised by the hubble length $cH^{-1}$ at the end of
inflation (see subsect. A), 
$\ln\left(0.002\,{\rm Mpc}^{-1}c/(a_fH_f)\right)=
-\ln(g_*^{1/2}T_{14})-53.6$, where $T_{14}\times 10^{14}GeV$ is the temperature
at the end of inflation.}
\end{figure}

\begin{figure}
\psfrag{PR}{${\cal P}_{\cal R}$}
\scalebox{0.7}{\includegraphics{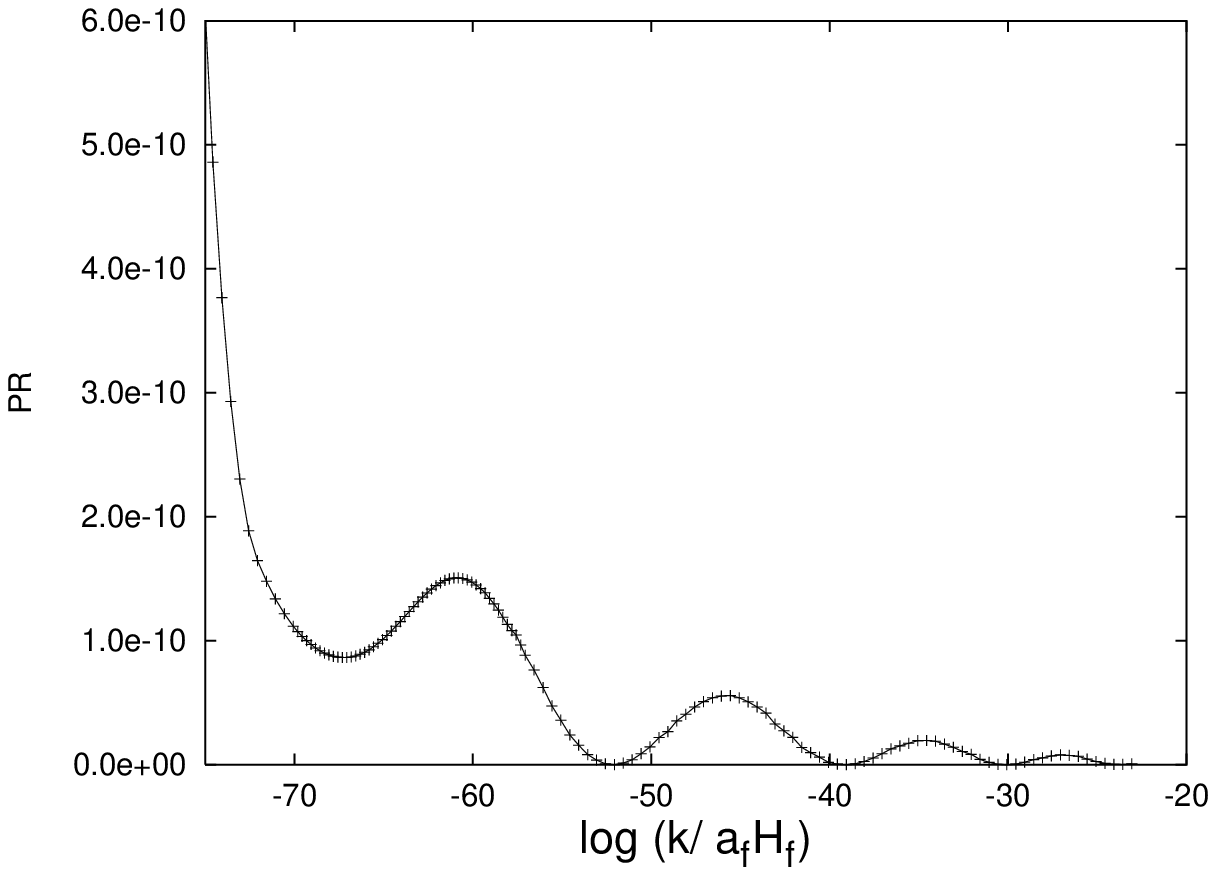}}
\scalebox{0.7}{\includegraphics{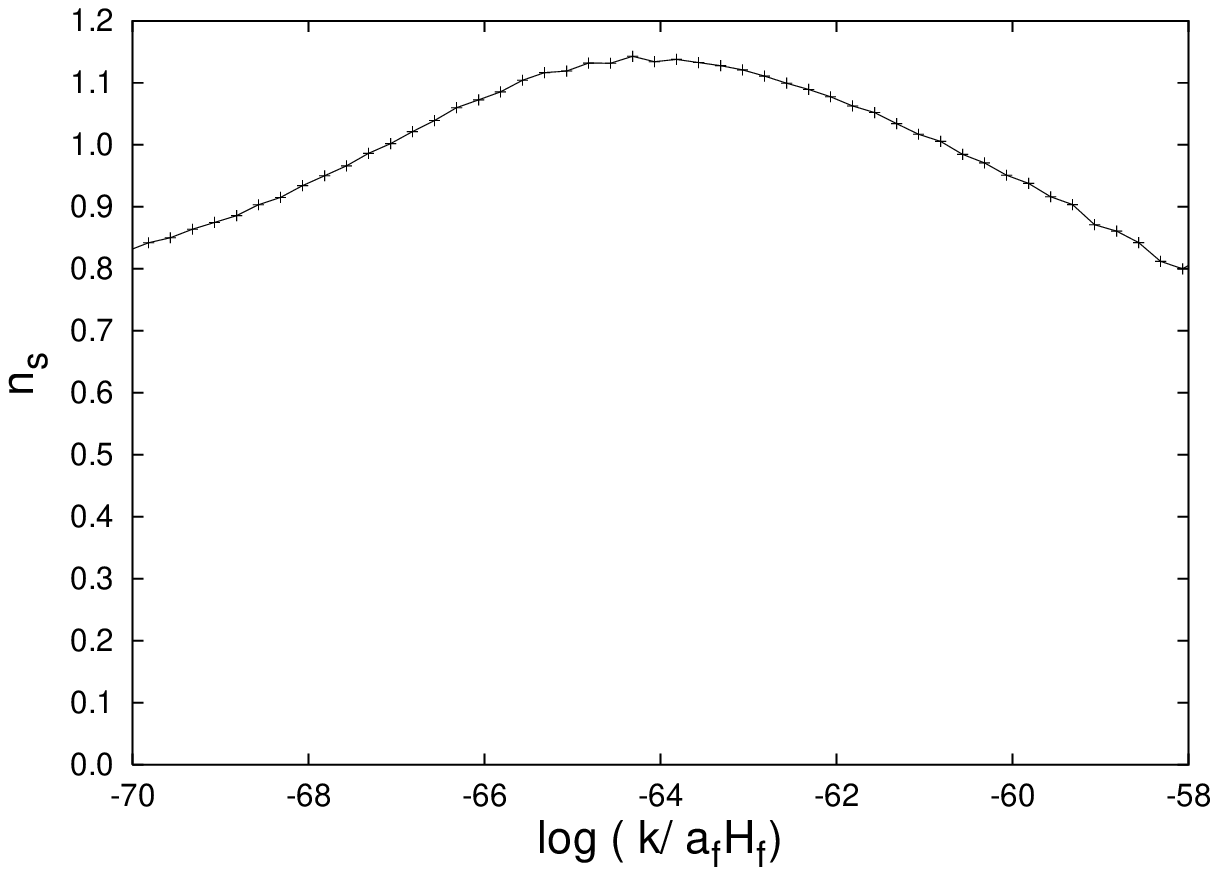}}
\caption{\label{fig5}The scalar fluctuation spectrum for the potential
$\frac14\lambda(\phi^2-\phi_0^2)^2$ with Hosoya's damping term. The spectral
index over a limited range of $k$ is also shown.}
\end{figure}

The spectral index for the first two models is plotted in figure \ref{fig4}.
The results are in close agreement with the analytical formula (\ref{speci}).
When the damping depends on the temperature, the spectrum is rather more
complicated and has the oscillatory form shown in figure \ref{fig5}. The
location of the zeros in this spectrum depend on the parameters $\Gamma_0$,
$\lambda$ and $\phi_0$. At an extremum in the spectrum, the index changes from
blue ($n_s>1$) to red ($n_s<1$) with increasing wavenumber $k$.

The observational situation is developing rapidly. The first year of cosmic
microwave background anisotropy measurements made by the WMAP satellite
suggest a spectral index $n_s<1$ \cite{map03}. However, the combination of the
microwave data with large scale structure data \cite{map03-2} produce a quite
different conclusion, that the spectral index runs from blue ($n_s>1$) to red
($n_s<1$) with increasing wavenumber $k$. Peiris et al \cite{map03-2} give
their best estimate of the spectral index as $n_s=1.2$ and $d n_s/d\ln
k=-0.077$ at $k=0.002 Mpc^{-1}$. 

The possibility of a spectrum which runs from blue to red is particularly
interesting because it is not commonly seen in inflationary models, which
typically predict red spectra. Examples of inflation with blue spectra do
exist however. The hybrid inflation models, which have two scalar fields, are
examples \cite{copeland94,linde97}. 

The numerical results show clearly that warm inflation can produce a spectrum
of density perturbations with blue runs to red behaviour. This behaviour is
seen quite generally for potentials with spontaneous symmetry breaking when
$\Gamma$ has the form Eq. (\ref{friction}) with $b=2$ and $c=0$, and is a
possibility when $b=2$ and $c=-1$.

\section{Conclusion}

The numerical results presented here show some of the possibilities which we
might expect from scalar density fluctuations in the warm inflationary
scenario. We have focused on the effects that result from various forms of the
damping coefficient $\Gamma$. Our main results are: 
\begin{itemize}

\item The non-stochastic evolution begins at the freezeout time $t_F$ when
the fluctuations in the inflaton are given by equations (\ref{dphi}) and
(\ref{ddphi}).

\item The spectral index when $\Gamma\equiv\Gamma(\phi)$ is given by three
slow-roll parameters Eq. (\ref{speci}). In this case warm inflation introduces
only
one extra slow-roll parameter.

\item The combination of `spontaneous symmetry breaking' potentials and
$\Gamma\equiv\Gamma(\phi)$ typically leads to $n_s>1$ on long scales and
$n_s<1$ on short scales.
 
\end{itemize}

Emphasis has been placed here on methods for solving the perturbation equations
rather than constructing a realistic inflaton potential. However, the
equations and the numerical integration have been set up in a way which makes
them easily adaptable to different models. The inflaton potentials may include
thermal corrections and different forms of the damping terms may also be taken
into account, for example using the friction coefficients calculated in
reference \cite{berera98,berera01,berera03}.

A better understanding of non-equilibrium thermodynamics will enable us to
calculate the friction term in the inflaton equation for a wider range of
conditions than is possible at present. Models of warm inflation are
restricted by consistency requirements \cite{berera98,yokoyama99,berera00},
many of which are a result of assumptions made due to the difficulties in
calculating far from equilibrium effects. Our results suggest that warm
inflation can easily produce a spectrum of density fluctuations that fits the
observational data and that this is a direction worth pursuing.

\bibliography{warminf.bib,inflation.bib,books.bib}

\end{document}